\documentclass[10pt,twocolumn,letterpaper]{article}

\usepackage{cvm}
\usepackage{times}
\usepackage{epsfig}
\usepackage{graphicx}
\usepackage{amsmath}
\usepackage{amssymb}

\usepackage{mathtools}
\usepackage{multirow}
\usepackage{colortbl}		%颜色，使用cellcolor\rowcolor等。'!'用于调整颜色深浅
\usepackage{makecell}

\usepackage[pagebackref=true,breaklinks=true,letterpaper=true,colorlinks,bookmarks=false]{hyperref}

\newcommand{\keywords}[1]{{\bf \emph{Keywords: #1}}}

\cvmfinalcopy % *** Uncomment this line for the camera-ready version

\def\cvmPaperID{353} % *** Enter the cvm Paper ID here

\ifcvmfinal\fi

\newcommand{\Skip}[1] {
}

\newcommand{\refFig}[1]{Figure \ref{#1}}
\newcommand{\refEq}[1]{Equation (\ref{#1})}

\newcommand{\refSec}[1]{Section \ref{#1}}
\newcommand{\refTab}[1]{Table \ref{#1}}

\begin{document}

%%%%%%%%% TITLE
\title{LDM: Large Tensorial SDF Model for Textured Mesh Generation}

% \and
% {\small\url{http://www.author.org/~second}}
% }

\author{%
Rengan Xie$^{1}$ \quad Wenting Zheng$^{1}$ \quad Kai Huang$^4$ \quad Yizheng Chen$^2$ \quad Qi Wang$^1$ \\
Qi Ye$^3$ \quad Wei Chen$^1$ \quad Yuchi Huo$^{1,2*}$ \\
$^1$State Key Lab of CAD\&CG, Zhejiang University \\
$^2$Zhejiang lab \\
$^3$State Key Laboratory of Industrial Control Technology, Zhejiang University\\
$^4$Institute of Computing Technology, Chinese Academy of Sciences\\
\texttt{\{rgxie,qi.ye\}@zju.edu.cn}, \texttt{\{wtzheng,chenwei\}@cad.zju.edu.cn}\\
\texttt{huangkai21@mails.ucas.ac.cn}, \texttt{chenyizheng123@126.com},\\
\texttt{\{wqnina1995, huo.yuchi.sc\}@gmail.com}
}

\maketitle
% \thispagestyle{empty}

%%%%%%%%% ABSTRACT
\begin{abstract}
Previous efforts have managed to generate production-ready 3D assets from text or images. However, these methods primarily employ NeRF or 3D Gaussian representations, which are not adept at producing smooth, high-quality geometries required by modern rendering pipelines. In this paper, we propose LDM, a novel feed-forward framework capable of generating high-fidelity, illumination-decoupled textured mesh from a single image or text prompts. We firstly utilize a multi-view diffusion model to generate sparse multi-view inputs from single images or text prompts, and then a transformer-based model is trained to predict a tensorial SDF field from these sparse multi-view image inputs. Finally, we employ a gradient-based mesh optimization layer to refine this model, enabling it to produce an SDF field from which high-quality textured meshes can be extracted. Extensive experiments demonstrate that our method can generate diverse, high-quality 3D mesh assets with corresponding decomposed RGB textures within seconds. The project code is available at \href{https://github.com/rgxie/LDM}{\textcolor{cyan}{https://github.com/rgxie/LDM}}.

\Skip{a novel model for lifting a single image to 3d content stepwise. In order to harness more information for novel view synthesis from progressively increasing multi-view image constraints, we present a novel diffusion model that generates novel images with the condition of a flexible number of images. Moreover, we propose a method that utilizes posterior knowledge, guiding the diffusion model to produce consistent images. This method generates several predictive results for a new viewpoint, considering constraints from neighboring images. Then, a specially designed evaluation strategy is employed to select the most compatible outcome with existing viewpoints. To reconstruct the 3D content from predicted images, we employ a view augmentation fusion strategy that minimizes pixel-level loss in generated sparse views and semantic loss in augmented random views, thereby avoiding inconsistencies in the predicted images. Qualitative and quantitative experiments show that StepwiseDream signiﬁcantly outperforms state-of-the-art single-view 3D reconstruction. Specifically, StepwiseDream generates highly consistent images and reconstructs 3d content with view-consistent geometry and detailed textures.}
    
\end{abstract}

\keywords{3D Generation model, Diffusion, Intrinsic decomposition, Relighting.}

%------------main body----------------

\section{Introduction}
\label{sec:intro}
Generating 3D content is vital for diverse applications and tasks. Recent 3D generation works leverage large 2D content generation diffusion models ~\cite{rombach2022high} to generate multi-view consistent images based on certain conditions, and elevate the 2D prior knowledge within diffusion models to 3d space through score distillation sampling (SDS)~\cite{poole2022dreamfusion,lin2023magic3d} or reconstruction methods~\cite{liu2023zero,liu2023syncdreamer}. Despite the impressive results, these require additional optimization time during the forward generation of 3D assets, often taking tens of minutes.

\begin{figure*}[ht] 
  \includegraphics[width=\textwidth]{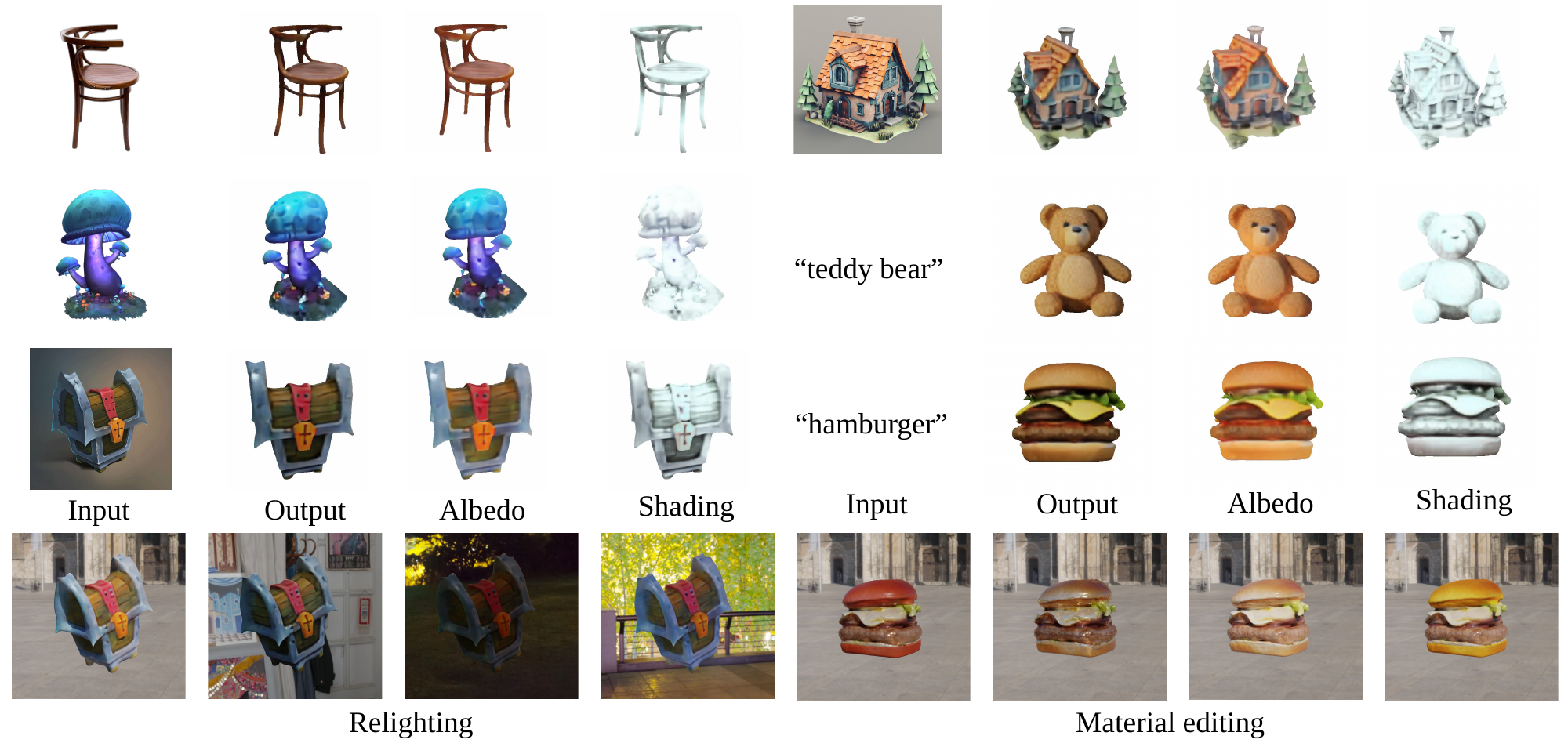}
  \caption{Given a text prompt or a single image, our framework can generate corresponding high-quality 3D assets within seconds, including illumination-decoupled texture maps,  facilitating integration into various applications, such as relighting and material editing.}
  \label{fig:teaser}

  \vspace{-0.3cm}
\end{figure*}

LRM ~\cite{hong2023lrm}, on the other hand, propose an end-to-end 3D Large Reconstruction Model that predicts the 3D model of an object from a single input image in about 5 seconds, requiring no optimization time during the forward process. LRM utilizes a Transformer to encode the input image as a condition and predicts a NeRF of the object in tri-plane representation. Instant3d ~\cite{instant3d} enhances LRM by using multi-view diffusion to generate multiple viewpoints from a single image. These views, encoded by an image encoder, serve as constraints to predict the target NeRF of the object, enabling the generation of higher-quality 3D assets from various perspectives. However, constrained by the overhead of volume rendering, LRM-based models can only be optimized under lower-resolution image patches. To achieve greater efficiency and higher resolution supervision, LGM and TMGS ~\cite{LGM, TMGS} utilize 3DGS~\cite{3dgs} as the representation of objects, employing UNet and Transformer architectures to predict and generate the Gaussian fields of objects. However, both the NeRF and 3DGS representations struggle with issues of unsmooth geometries and suboptimal geometric quality.

SDF representation has been proven capable of reconstructing high-quality. However, introducing SDF to the learning based 3D generation pipeline encounters the challenges of high memory computation and unstable convergence. Zero12345 and Zero12345++ ~\cite{zero12345, zero12345plus} introduce 3D CNNs or 3D diffusion to predict SDF volumes from image inputs. Nevertheless, the quality of the output is constrained by the resolution of the volume grid, and both 3D diffusion and 3D CNNs incur significant computational overhead. As observed in CRM ~\cite{crm}, predicting the correct geometry using pure RGB images is extremely hard, and it adds extra geometry information into the inputs of the pipeline to help the convergence of geometry generation with the SDF representation.

In addition to the geometry component, generating lighting-decomposed texture maps is crucial for editing and relighting 3D assets in downstream applications. However, existing learning-based 3D generation methods cannot provide lighting-decomposed texture maps, their textures embed shading effects such as shadows and highlights.

To generate 3D assets with high-quality meshes and illumination-decoupled textures within the end-to-end learning based framework, we propose LDM, a novel 3D generation pipeline with tensorial SDF representation and decoupled color field, as shown in~\refFig{fig:teaser}. Specifically, to improve the object surface quality and reduce memory requirements, we combine an SDF representation for the geometry with a low-rank tensorial representation in TensoRF\cite{chen2022tensorf}. Furthermore, to adapt to the light decoupling task in the generative approach, we introduce a shading model that separates the final output into an albedo color and a shading color. 

To predict a decoupled representation, SDF values, albedo color, and shading color at any point in space are decoded from the tensorial representation and supervised by reference albedo and color images using volume rendering. However, the method used to convert SDF to a density field in reconstruction works, such as VolSDF~\cite{yariv2020multiview}, is not applicable to our generative approach.  In these works, a parameter $\beta$ is required to control the sharpness of the conversion and must be adjusted based on the convergence status of the SDF for each scene, while a large $\beta$ means smoother geometry. Unlike reconstruction tasks that focus on a single object or scene, the generative approach needs to handle a variety of geometries, some smooth and others highly detailed. To address this issue, we introduce an adaptive $\beta$ adjusting schedule, making $\beta$  an optimizable parameter to allow $\beta$ to be adjusted directly by the gradient descent from the final losses. 

To achieve finer texture and geometry details, a differentiable iso-surface extraction layer like Flexicube~\cite{flexible} can be integrated to enable high-resolution image supervision following the tensorial SDF representation. Specifically, directly training to predict the SDF field using the Flexicube~\cite{flexible} rendering layer fails to converge due to sparse view supervision for each object and gradient descent impacting only the vicinity of the SDF grid vertices. To address this, a preliminary volume rendering training phase is essential. This phase samples points throughout the entire space to stabilize training and provide an initial SDF, allowing the Flexicube renderer to refine local quality. Extensive experiments show that our method is capable of generating diverse, high-quality 3D mesh assets along with decomposed RGB textures in just seconds.

In summary, our main contributions are: 
\begin{itemize}
    \item We propose the first feed-forward framework capable of generating high-quality meshes with illumination-decoupled RGB textures from text or a single image input in just a few seconds.
    \item We introduce tensorial representation into the generation of objects, accurately representing illumination-decoupled SDF fields, and enhancing convergence speed. An adaptive conversion of SDF to density strategy is proposed to enable the convergence of the novel representation in our feed-forward 3D generation framework.
    \item We integrate a gradient-based mesh optimization layer to train our framework, enabling it to produce an SDF field from which high-quality triangular meshes and illumination-decoupled textures can be extracted.

\end{itemize}

\section{Related Work}
\label{sec:relatedwork}
% % at most 1 page with a paragraph for the brief contributions
%1st draft

\paragraph{Diffusion Models for Multi-view Synthesis} 
% \todo{Diffusion models for multi-view synthesis.} 
Zero123~\cite{liu2023zero} demonstrates that large diffusion models have learned rich 3D priors about the visual world, even though they are only trained on 2D images. Furthermore, they introduce the first refined diffusion model capable of generating plausible images of objects from any viewpoint. A series of subsequent works~\cite{shi2023zero123++,shi2023mvdream,liu2023syncdreamer,long2023wonder3d} have focused on enhancing the 3D consistency and resolution of multi-view images generated by diffusion models.  Zero123++ ~\cite{shi2023zero123++} proposes that generating consistent multi-view images hinges on accurately modeling their joint distribution and integrates six images in a 3×2 layout into a single frame for simultaneous multi-view generation during one diffusion process.  MVDream~\cite{shi2023mvdream} and Wonder3D~\cite{long2023wonder3d} both recommend implementing multi-view attentions to enable feature sharing during the multi-view generation process. Free3D~\cite{free3d} introduces Plücker Embeddings of pixels in images' corresponding rays and a Ray Conditional Normalization layer during the diffusion generation process, which helps to produce multi-view images with more accurate perspectives. Although these methods have generated impressive multi-view images, achieving 3D consistency among them remains a challenge.

%stash for space
\Skip{
SyncDreamer~\cite{liu2023syncdreamer} unprojects features from multiple views into a single view frustum volume, from which a depth-wise attention layer in its diffusion model extracts information to enhance the 3D consistency of generated images. Unlike Zero123, which produces views independently, Zero123++ integrates six images in a 3×2 layout into a single frame for simultaneous multi-view generation during one diffusion process.ensuring greater consistency in the results. Moreover, beyond cross-attention between multiple views, 
 
}

% image base relighting(human+object) 
\paragraph{Lifting 2D Diffusion for 3D Generation} 

Recent breakthroughs in multi-view diffusion model ~\cite{rombach2022high} have rapidly sparked interest ~\cite{poole2022dreamfusion,liu2023zero} in reconstructing 3D content generation from various conditions, such as Text and images, known as 2D-lifting, include distillation-based approaches and reconstructed-based approaches. \textit{Distillation-based approaches}, exemplified by Dreamfusion and Magic3D ~\cite{poole2022dreamfusion,lin2023magic3d}, leverage 2D diffusion models for optimizing 3D representations through score distillation sampling (SDS). \textit{The reconstructed-based approaches} are notably represented by Zero123 and SyncDreamer~\cite{liu2023zero,liu2023syncdreamer}, which involves generating images with multi-view consistency using diffusion models, followed by employing volume rendering techniques to optimize 3D representations, such as neural radiance fields (NeRF) ~\cite{mildenhall2020nerf}or signed distance functions (SDF)~\cite{wang2021neus}. These methods have achieved impressive results, but there are still some issues. The distillation approach tends to struggle with multi-faceted problems and requires a longer time to refine a single shape. While Zero123~\cite{liu2023zero} is capable of producing images from any viewpoint using a single input image and its relative pose, it faces challenges in maintaining consistency across multi-view images, leading to 3D representations with blurred geometric details. Moreover, these methods require additional optimization time during the forward generation of 3D assets.

\paragraph{Feed-forward 3D Generative Models}

Currently, there are some feed-forward methods capable of generating 3D assets directly from text and image inputs without optimization. LRM~\cite{hong2023lrm} introduced the first end-to-end 3D Large Reconstruction Model, predicting a NeRF of the object from a single image in just 5 seconds. Since LRM only uses a single image as the condition for the transformer, the resulting object's back side exhibits sparse coloring, inconsistent with the front. Instant3D~\cite{instant3d} improves LRM by generating multiple viewpoints from a single image through multi-view diffusion. The encoded views act as constraints to predict the object's NeRF, enhancing the quality of 3D assets from different perspectives. 3DGS~\cite{3dgs} has emerged as a new attempt at object representation because it can quickly produce high-quality, high-resolution images from any viewpoint through splatting and rasterization. LGM~\cite{LGM}encodes attributes of 3D Gaussians from multiple views into splatter images~\cite{splatter} and uses a UNet to predict these images from RGB inputs.TMGS~\cite{TMGS} uses a transformer to predict tri-plane attribute fields, enriching point cloud models of objects with additional properties to achieve a 3DGS representation.
Nevertheless, both 3DGS and NeRF representations encounter difficulties with issues of uneven geometries and suboptimal geometric quality. SDF representation has been proven capable of reconstructing high-quality geometries Zero12345 and Zero12345++ ~\cite{zero12345, zero12345plus} use 3D CNNs or 3D diffusion to predict SDF volumes from images, but the output quality is limited by the volume grid's resolution. CRM~\cite{crm} uses diffusion to predict six Canonical Coordinates Maps from a single input and optimizes a UNet to predict a tri-plane from these CCMS, which includes an SDF field. 
CRM integrates a differentiable iso-surface extraction layer~\cite{flexible} to optimize the generated SDF. Concurrent work InstantMesh~\cite{instantmesh} utilizes an LRM-based model to predict 3D assets represented as NeRF from multi-view images produced by multi-view diffusion. These NeRFs are subsequently refined into a Signed Distance Field (SDF). These efforts have produced impressive textured meshes; however, their textures embed shading effects such as shadows and highlights, making these meshes unsuitable for relighting and material editing. This limitation hinders their use in downstream applications.

%stash for space 
 \Skip{
The encoded views act as constraints to predict the object's NeRF, enhancing the quality of 3D assets from different perspectives. NeRF representations are limited by the memory overhead of volume rendering, preventing high-resolution modeling. each pixel holding 14-dimensional properties like RGB and rotation. and the combination results in the 3DGS representation of the object., allowing for smoother triangular meshes that enhance downstream applications.
 , enabling mesh extraction and the rendering of high-resolution results. 
 }

\section{Method}
\label{sec:overview}

\begin{figure*}[ht] 
\centering 
\includegraphics[width=\linewidth]{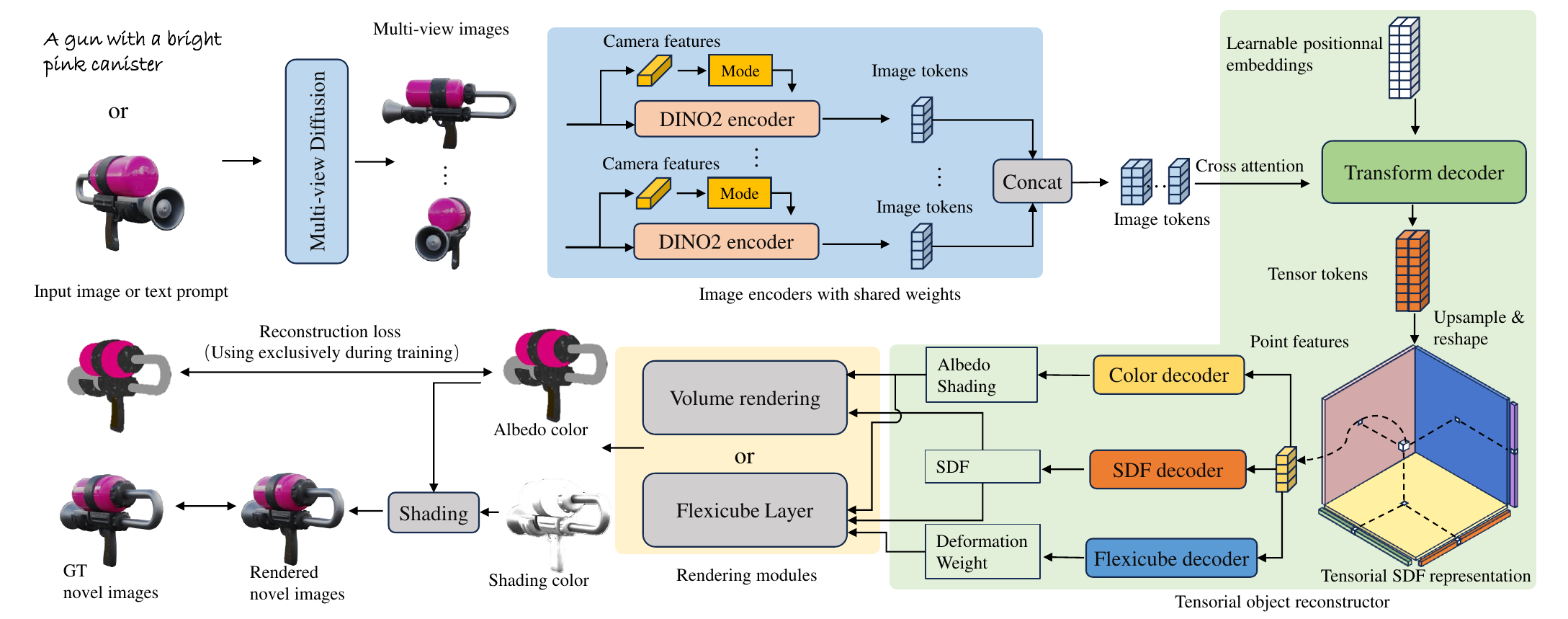}
\caption{\textbf{The overview of our framework.} When given an image or text prompt condition, we first utilize a diffusion model to generate multiple viewpoint images. These images are then encoded into image feature tokens using the DINO2 image encoder. Subsequently, these tokens are fed into a transform-based tensorial object reconstructor, resulting in a tensorial SDF representation. The tensorial SDF representation can be further rendered using volume rendering or the Flexicube render layer to produce images or extract meshes.}

\label{fig:overview} 
\vspace{-0.2cm}
\end{figure*}

We propose the LDM, a framework as shown in~\refFig{fig:overview} that takes a single image or a text prompt as input and generates a corresponding triangular mesh equipped with illumination-decoupled textures. We first employ a multi-view diffusion model to generate multiple viewpoint images of the target object, conditioned on either an input text prompt or image, as detailed in~\refSec{sec:mvdiffusion}. In~\refSec{sec:tensorialsdf}, we discuss the use of tensorial SDF representation for generating objects, which offers robust expressive power. Next, we introduce our two-stage training strategy. In~\refSec{sec:rec}, we introduce how to train a transformer-based model to generate tensorial SDF representations from sparse multiple views using volume rendering in the first stage. Finally, in the second stage detailed in ~\refSec{sec:mccube}, we introduce a gradient-based mesh optimization layer, Flexicube~\cite{flexible}, to refine the generative model and enhance the quality of the extracted textures.

\subsection{Conditional Multi-view Generation}
\label{sec:mvdiffusion}

% How to train a diffusion model support to predict novel views with the input of multi-view conditions and progressively synthesize multi-view images.

Since Zero123~\cite{liu2023zero}, numerous diffusion models capable of producing conditional multi-view images have emerged, such as MVdream, Imagedream, SyncDreamer, and Zero123Plus~\cite{shi2023mvdream,imagedream,liu2023syncdreamer,shi2023zero123++}. We utilize MVDream for text inputs and ImageDream for image inputs. Both models are engineered to generate multi-view images from four orthogonal azimuths at a fixed elevation, which will serve as input conditions for the subsequent generation pipeline. Compared to other methods, MVDream and ImageDream employ joint training with synthetic data from the Objaverse dataset~\cite{objaverse} and real data from the large-scale text-to-image (t2i) dataset, LAION5B~\cite{laion}. Thus, We choose these models in our framework, aiming to boost our generalization capabilities for realistic object generation.

Although the implementation of a cross-view attention mechanism in those models has improved the consistency of the generated multi-view images, yielding impressive results, subtle 3D inconsistencies may still occur. To simulate this perturbation, inspired by previous methods~\cite{LGM,hong2023lrm}, we randomly apply a grid distortion to the input views and introduce a camera jitter to the camera poses during training. By deliberately introducing these inconsistencies during training, our framework is better equipped to handle the 3D inconsistencies encountered in images generated through multi-view diffusion.

\subsection{Tensorial SDF and Decoupled Color Field}
\label{sec:tensorialsdf}

Recently, many studies have proposed various implicit neural representations for 3d objects or scenes, combined with differentiable rendering techniques, achieving impressive reconstruction results. NeRF~\cite{mildenhall2020nerf} employs an MLP to encode the radiance field of a scene object and uses volume rendering to transform these volume densities and colors within the radiance field into an image from the target viewpoint. Unlike NeRF, TensoRF~\cite{chen2022tensorf} represents the scene's radiance field with a 4D tensor. This tensor describes a 3D voxel grid, each voxel containing multi-channel features. TensoRF further decomposes this 4D tensor into several compact, low-rank components, greatly reducing memory requirements and improving rendering quality. NeRF-based methods often suffer from the issue of uneven geometries in reconstructed object surfaces. To address this, works such as VolSDF and NeuS~\cite{yariv2021volume,wang2021neus} propose replacing density with SDF in implicit representations to enhance the geometric quality of the reconstructions, resulting in smoother surfaces. Furthermore, TensoSDF~\cite{li2024tensosdf}, which combines tensorial representation with SDF, has proven effective in encoding various material features, such as albedo and metallic properties, in inverse rendering tasks. Inspired by these studies, we propose employing a tensorial SDF representation to depict both the illumination-decoupled color field and SDF field of the objects generated in this work.

\textbf{\textit{Tensorial representation.} }
We utilize the Vector-Matrix factorization method proposed by TensoRF as our tensorial representation. Specifically, our tensorial representation is formulated as follows:

\begin{equation}
V_p = V^X_k \circ M^{YZ}_k\oplus V^Y_k \circ M^{ZX}_k  \oplus V^Z_k \circ M^{XY}_k ,
\end{equation}
where $V_p$ denoted the feature vector of position $p$. In addition, $V^m_k$ and ${M}_k^{\tilde{m}}$ represent the $k$-th vector and matrix factors of their corresponding spatial axes $m$, and $\tilde{m}$ is the two axes orthogonal to m (e.g., $\tilde{X}=YZ$). $\circ$ and $\oplus$ denote the element-wise
multiplication and concatenation operations. Moreover, unlike TensoRF, which uses two separate tensor fields to encode geometry and appearance, we leverage a single shared tensorial field for both. This approach enhances the correlation between geometry and appearance, as mentioned in~\cite{li2024tensosdf}.

\textbf{\textit{Model SDF and decoupled color field.}}
Our shading model differs from traditional PBR-based shading formulations and is more akin to intrinsic image decomposition works~\cite{li2018learning,careaga2023intrinsic,janner2017self}, where the color is divided into a reflectance component (albedo) and a shading component that captures illumination information in image space. This approximation has been proven to be a fast and effective approach. Using PBR-based shading models requires modeling and predicting the ambient lighting of the input condition images when generating 3D assets, which could lead to unstable model training and difficulty in convergence. Additionally, the condition images are generated by a diffusion model, they cannot ensure consistent lighting across multiple images, which further complicates the direct prediction of lighting. Adopting a simplified shading method that is closer to intrinsic image decomposition is a trade-off approach for use in feed-forward generative models.

Therefore, we decompose the final rendering color into two parts. The first part is the albedo color, which reflects the inherent color of the object and remains constant for each object. The second part is the shading color, which is determined by the varying lighting conditions the object receives and its different material properties. Given the decomposed appearance field and SDF field, we introduce multiple MLP decoders to decode the features extracted from the tensorial representation. This approach enables predicting various material properties and SDF from compact features. Formally, 

\begin{equation}
s = \Theta_s(V_p, p),  (c_a,c_r) = \Theta_c(V_p, p),  
\end{equation}
where $\Theta_s,\Theta_c$ represent the MLPs used to decode the SDF value, albedo color, and shading color. Equipped with the albedo color $c_a$ and shading colors $c_s$, the final rendering color $c$ of point $p$ can be calculated as, %$c = c_a \cdot c_r$. 
\begin{equation}
c = c_a \cdot c_r  .
\end{equation}
Once the albedo map with decoupled illumination is obtained, artists can enhance the 3D assets with PBR materials such as metallic and roughness, tailoring them to specific use cases and making them ready for application. In new scenes, the shading color can be calculated based on different materials and lighting conditions, following the rendering equation~\cite{kajiya1986rendering}.

\textbf{\textit{Adaptive conversion of SDF to density.} }
Currently, we can obtain the SDF of any point on the object, but SDF cannot be directly used for volume rendering. Follow VolSDF~\cite{yariv2021volume}, we model the conversion process from SDF value $s$ to the corresponding density $\sigma$ by the following formula:
\begin{equation}
\label{eq:sigma2sdf}
\sigma=\begin{dcases}
\frac{1}{\beta}\Bigl(1-\frac{1}{2}exp\Bigl(\frac{s}{\beta}\Bigr)\Bigr)& \text{if}\ s<0, \\ 
\frac{1}{2\beta} exp\Bigl(-\frac{s}{\beta}\Bigr)\Bigr)& \text{if}\ s \geq   0, \\
\end{dcases}
\end{equation}
where $\beta > 0$ is an important parameter that controls the sharpness of the conversion from SDF to density. A large beta will result in the density field obtained from the SDF conversion being too smooth, causing the loss of geometric details. Conversely, a small beta will lead to geometric fragmentation.
% \begin{wrapfigure}{r}{0.50\textwidth}

% \includegraphics[width=1\linewidth]{figures/beta_schedules.png}
% \caption{Comparing model training performance across different Beta schedules.}

% \label{fig:beta_schedules} 

% \end{wrapfigure}

\begin{figure}[ht] 
\centering 
\vspace{-0.2cm}
\includegraphics[width=1\linewidth]{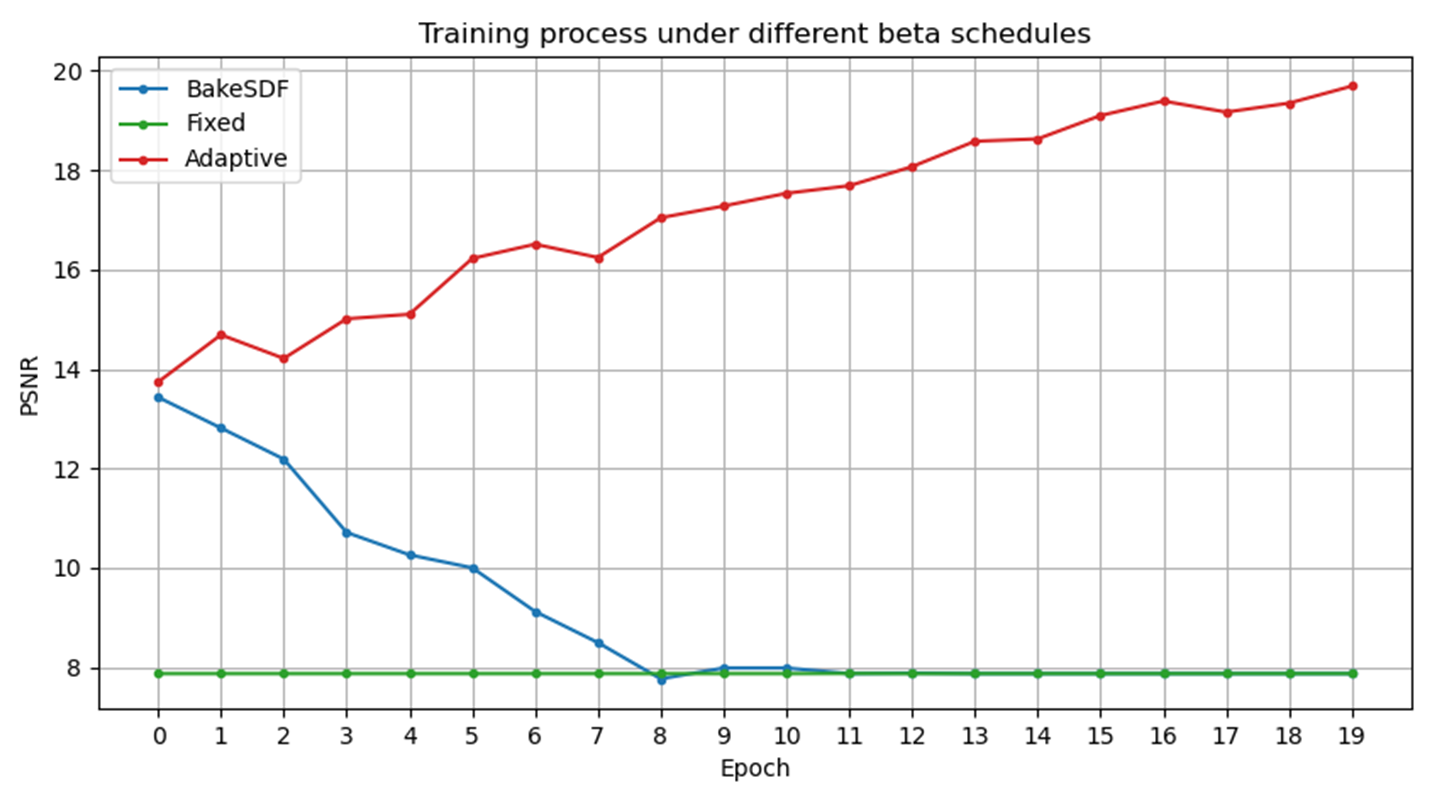}
\caption{Comparing model training performance across different Beta schedules.}

\label{fig:beta_schedules} 
\vspace{-0.3cm}
\end{figure}

To determine the appropriate beta value, several schemes are conducted.
The first scheme is using the empirical mean of beta values from the reconstructed models. We reconstructed 200 scenes from multi-view images using VolSDF~\cite{yariv2020multiview} and collected the optimized beta values from the 200 VolSDF models. The mean of these values sets the beta value to train our network. However, starting with a too-small beta made the network ultimately fail to converge. 
The second scheme is a linear scheduling mechanism for beta to increase training stability, following the approach of BakeSDF~\cite{yariv2023bakedsdf}. We initialize beta with a fixed value and gradually decrease it as the training epochs progressed, as $\beta=\beta_0 \left( 1 + \frac{\beta_0 - \beta_1}{\beta_1}t^{-1} \right)$, where $t$ goes from 0 to 1 during training, $\beta_0=0.1$
, and $\beta_1$ for the three ray points hierarchical sampling is 0.015, 0.003,
and 0.001, respectively. Unfortunately, this also failed. We attribute it to that the manually set beta decay strategy might not align with the convergence rate of our generation network. 
Therefore, we introduce an adaptive beta adjust schedule and allow the beta value to be adjusted directly by the gradient descent process instead of a manual schedule. Specifically, we initially set beta to a relatively large value of 0.1 and made it an optimizable parameter. We train it together with the image encoder and tensorial object reconstructor within the framework. We found that this approach allows beta to gradually decrease as the model acquired more knowledge, eventually converging with the model. As seen in~\refFig{fig:beta_schedules}, this technique effectively guides the model toward convergence.

\subsection{Feed-forward Large Reconstruction Model} 
\label{sec:rec}

In this section, we describe how a transformer-based model reconstructs a tensorial representation from sparse view images generated by multi-view diffusion. Reconstructing 3D representations from sparse view inputs has always been a challenge in the community, with traditional methods struggling to incorporate and apply prior 3D knowledge. In recent years, diffusion models trained on vast datasets have demonstrated a robust understanding of prior knowledge, capable of generating images under various controlled conditions. This has sparked interest in exploring large models for 3D generation. Inspired by Instant3D and LRM ~\cite{instant3d,hong2023lrm}, we propose a new transformer-based model architecture that includes an image encoder and a tensorial object reconstructor to predict a tensorial SDF field conditioned on multi-view image features

Specifically, we use a vision transformer, DINO2~\cite{dino}, to encode feature tokens $F_I$ from multi-view images as shown in~\refFig{fig:overview}. To ensure that the resulting feature tokens incorporate multi-view camera information, we modulate the camera information as described in LRM~\cite{hong2023lrm} and inject it into the image encoder using AdaLN~\cite{huang2017arbitrary,peebles2023scalable}. Furthermore, we arrange the tensor vectors $V^m_k$ and ${M}_k^{\tilde{m}}$ into learnable tokens and feed these tokens into a feature decoder composed of a sequence of transformer layers, resulting in a series of tensor tokens. It should be noted that the information from the conditioned image feature $F_I$ is connected to the cross-attention layer within these transformer layers. Finally, we reshape and upsample the tensor tokens into the final tensorial SDF representation. Then, all the transformer architecture can be trained in an end-to-end manner using image reconstruction loss at novel views images rendered from the tensorial SDF representation using volume rendering~\cite{mildenhall2020nerf}.  More details of the network structure can be found in the supplementary material. The training loss in this stage is defined as,

\begin{equation}
\begin{aligned}
    \mathcal{L}_{1} &= \mathcal{L}_{mse}(I_{rgb}, I^{GT}_{rgb})+\mathcal{L}_{mse}(I_{\alpha}, I^{GT}_{\alpha})  \\
     & + \mathcal{L}_{mse}(I_{h}, I^{GT}_{h})+ \lambda_{vgg} \mathcal{L}_{lpips}(I_{rgb}, I^{GT}_{rgb}),
\end{aligned}
\label{eq:l1}
\end{equation}
where $I_{rgb},I_{\alpha},I_{h}$ respectively represent the predicted final color image, albedo image, and mask image. Following LRM~\cite{hong2023lrm}, we apply mean square error loss and VGG-based LPIPS loss~\cite{zhang2018unreasonable} to the predicted images and corresponding reference images. During training, we set $\lambda_{\text{vgg}}=2$.

\Skip{One-2-345 and its upgraded version~\cite{zero12345,zero12345plus} respectively utilize 3D CNNs and 3D diffusion models to predict SDF volume grids from features extracted from multi-view images. Both methods have achieved commendable results, but they suffer from the substantial computational overhead of the 3D CNN architecture, which prevents the enhancement of the resolution of the SDF volume grids.}

\subsection{Lifting SDF to Fine Mesh} 
\label{sec:mccube}

\begin{figure*}[h] 
\centering 
\includegraphics[width=\linewidth]{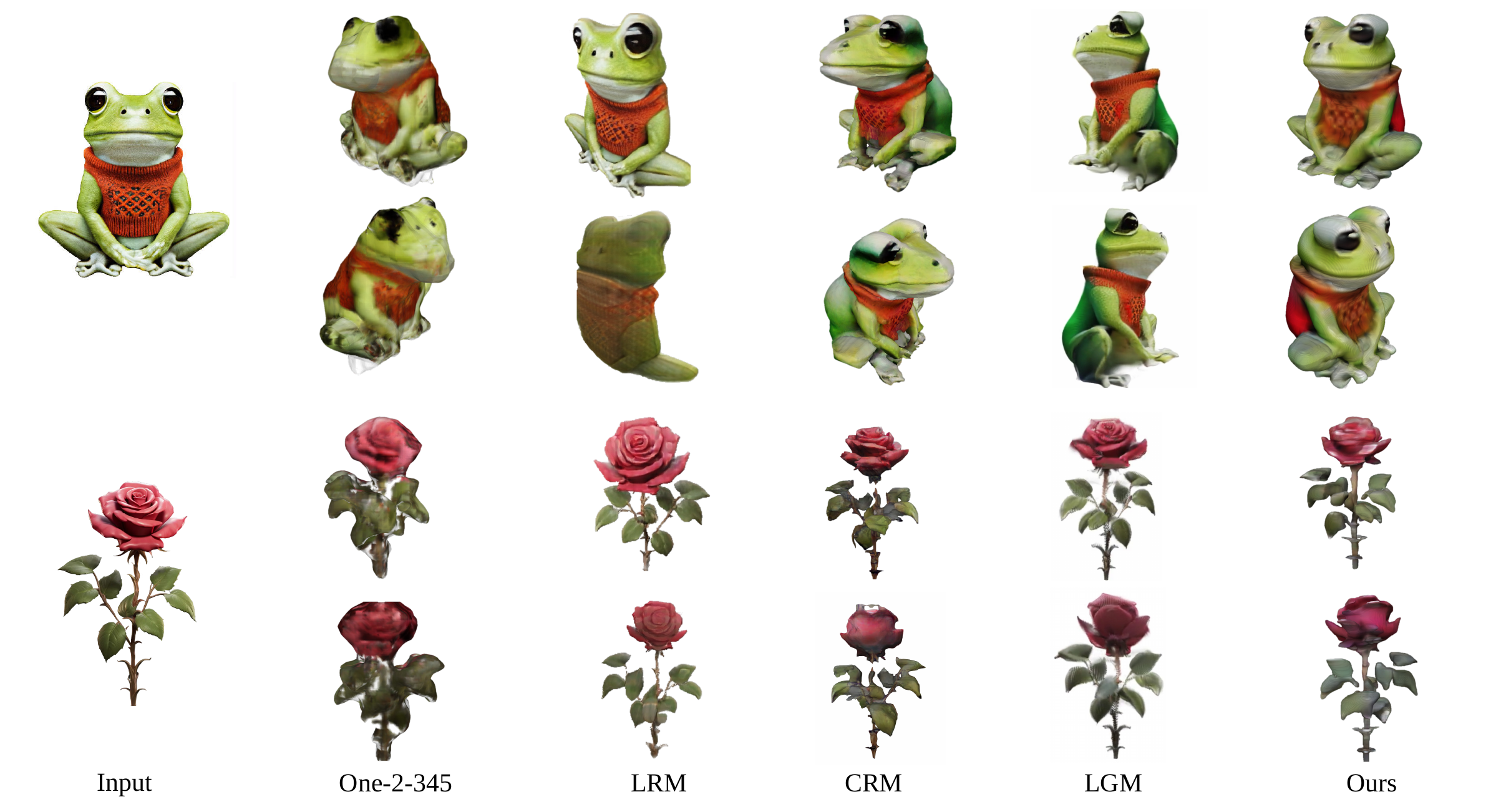}
\caption{Qualitative comparison with baselines shows that our method produces high-quality 3D assets with smooth geometry and clear textures, which align well with the input image.}
\vspace{-0.3cm}
\label{fig:compare_image} 
\end{figure*}

Till now, we have introduced a transformer structure that enables the generation of a tensorial SDF representation from sparse images. We can directly extract geometric surfaces from the SDF representation using Marching Cubes~\cite{mccube} after training. However, the strategy has the issue that given the high computational cost of volume rendering, only a patch of the image is rendered during training, thus being unable to utilize full-resolution supervisory images fully. In addition, a gradient-based mesh optimization layer after an SDF field can improve the silhouette quality as described in Nvdiffrec~\cite{nvdiffrec}. 

FlexiCubes~\cite{flexible} is a state-of-the-art differentiable iso-surface extraction layer. Employing the differentiable method to extract meshes from the tensorial SDF representation, rasterizing, and rendering at high resolution for supervised learning is beneficial for achieving finer details. One strategy is to use the Flexicube render layer from the outset to replace the computationally expensive volume rendering. However, due to the design of the discrete SDF grid, under sparse view supervision, each gradient descent step affects only the vicinity of the SDF grid vertex. Therefore, directly using Flexicube from the start during training cannot get the network to converge. CRM~\cite{crm} has similar observations:  predicting the correct geometry directly with Flexicube is extremely difficult using pure RGB images. In contrast, volume rendering sampling points throughout the entire space and gradients affect the entire space. 

\textbf{\textit{Two-stage training from global to local.} }
Therefore, we combine the advantages of both training with volume rendering and Flexicube and propose a two-stage training strategy. In the first stage, we use the more stable volume rendering for the learning of the global features of the network, allowing the network to achieve global convergence. In the second stage, we initialize the training with the model from the first stage and use Flexicube to optimize the local features, achieving higher resolution textures. During the second stage, to predict a set of weights in each grid cell and the deformation of each grid vertex following Flexicube~\cite{flexible}, we additionally introduced two MLP decoders to decode the tensor features from the tensorial SDF:

\begin{equation}
d = \Theta_d(V_p, p),  w = \Theta_d(V_p, p),
\end{equation}
where $\Theta_d,\Theta_w$ represent the MLPs used to decode the grid point deformation value and weights of the grid cell, which are needed for the FlexiCube rendering pipeline, respectively. The model is then trained under supervision using the following loss:

\begin{equation}
\begin{aligned}
\mathcal{L}_{2} &= \mathcal{L}_{1} + \lambda_{d} \left\| I_{d} - I_d^{gt} \right\|_1 
 + \lambda_{\text{reg}} \mathcal{L}_{\text{reg}} ,
\end{aligned}
\label{eq:l2}
\end{equation}
%+\lambda_{n} \left( 1 - I_{n} \cdot I_{n}^{gt} \right)
where $I_{d}$ represent the rendered depth map, $I_{d}^{gt}$ are the corresponding reference images. The $L_{reg}$ is the regularization term following FlexiCubes~\cite{flexible}. During training, we set $\lambda_{\text{d}}=0.5$, $\lambda_{\text{reg}}=0.005$.

\section{Experiment}
\label{sec:experi}

\begin{figure*}[ht] 
\centering 
\includegraphics[width=\linewidth]{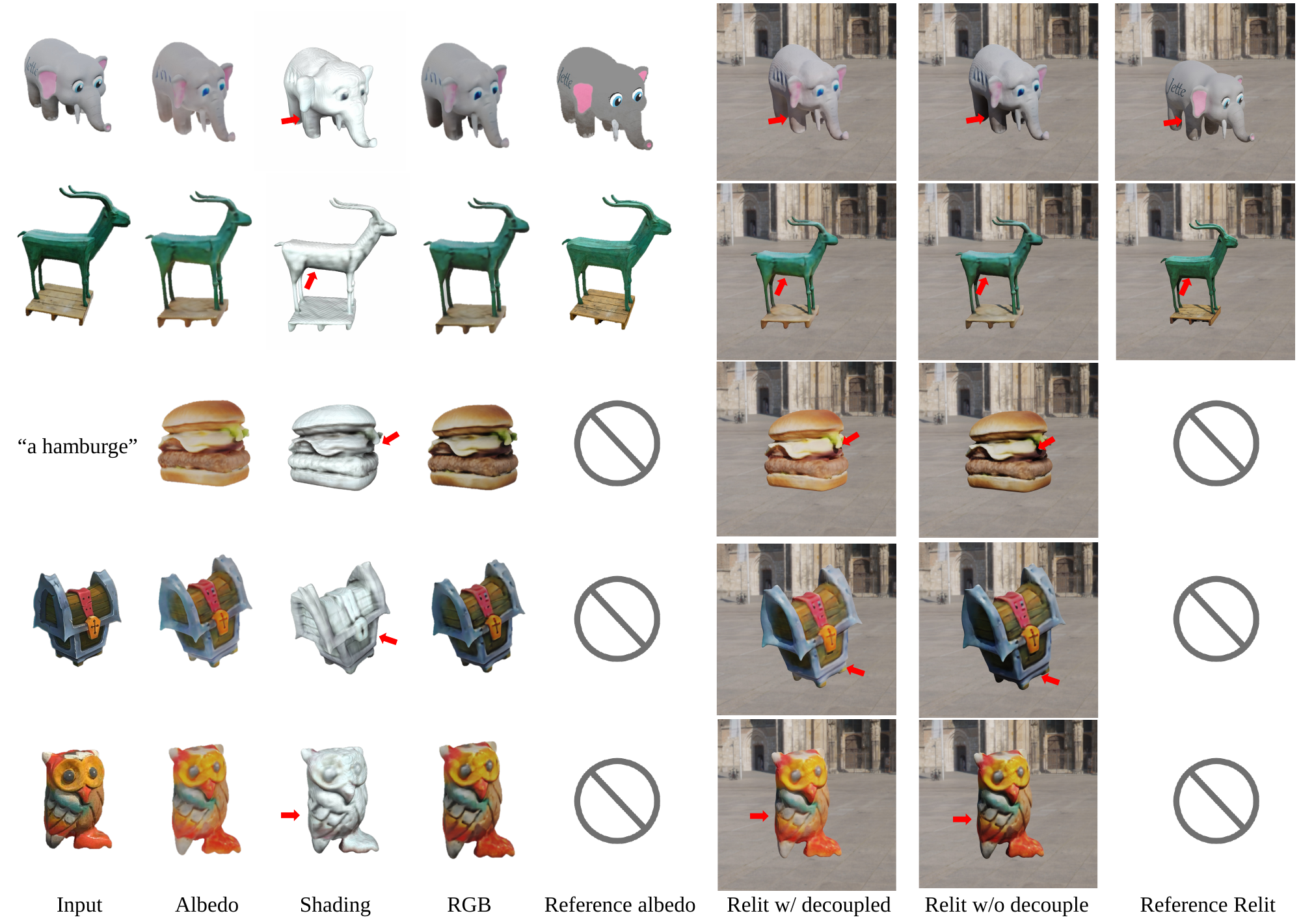}
\caption{The effect of illumination decoupled texture. We perform relighting in new scenes for both illumination-decomposed textures and non-decomposed textures. The 3D assets without illumination decomposition display incorrect shadows in the new scenes.}

\label{fig:ablation_decounpled} 
\end{figure*}

\subsection{Implementation Details}

\paragraph{Training Datasets}
We train the model in GObjaverse from RichDreamer~\cite{richdreamer}, which is rendered using the TIDE renderer on Objaverse~\cite{objaverse} and includes G-buffer rendering data such as albedo, RGB, depth, and normal map images from multiple views in resolution of 512$\times$512. Following LGM~\cite{LGM}, we utilize a filtered subset, which excludes low-quality 3D models, resulting in a final set of around 80K 3D objects. Specifically, this dataset includes 36 random views of a centered object, with elevations ranging from -5° to 30°, and two additional views for the top and bottom. Although the camera poses for images produced by multi-view diffusion are predetermined, during training, a random subset of 8 images is selected. 4 images are used as inputs to the model, and the remaining 4 are employed for novel view supervision. This strategy not only makes the model robust to inconsistencies in inference inputs but also ensures that the image encoder is sensitive to camera poses. For inference, images with fixed camera poses generated in the first stage are fed to the tensorial SDF reconstruction.

\begin{figure}[ht] 
\centering 

\includegraphics[width=1\linewidth]{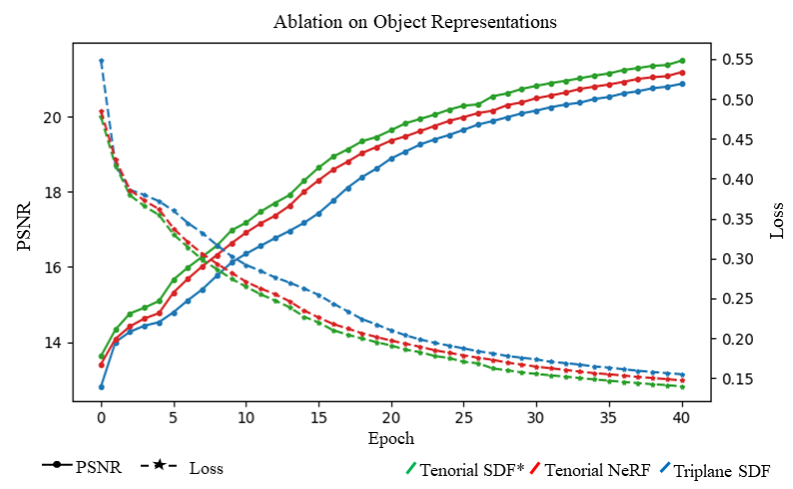}
\caption{Comparing model training performance across different object representations, Tensorial SDF emerges as the outstanding performer.}

\label{fig:ablation_reps} 
\vspace{-0.6cm}
\end{figure}

\begin{table}[h]
    \setlength{\tabcolsep}{2.2mm}
    \centering
    \vspace{-0.2cm}
    \caption{The features of different methods, our framework provides comprehensive feature support. }
    \vspace{0.2cm}
    \begin{tabular}{l|cccc}
    \Xhline{0.8pt}
    Method      & Tto3d & Ito3d & Rep. & Illum. dec. \\ \hline
    LRM         & $\times$               & \checkmark              & NeRF                & $\times$                                    \\
    LGM         & \checkmark                 &\checkmark             & 3DGS                  & $\times$                                     \\
    CRM         & $\times$                     & \checkmark                   & SDF                        &  $\times$      \\   
    Instand3d   & \checkmark               & \checkmark                & NeRF                 &  $\times$                                  \\
    
    Ours        & \checkmark               & \checkmark                 & TensoSDF                   & \checkmark                      \\
    \Xhline{0.8pt}
    \end{tabular}
    \vspace{-0.5cm}
    \label{tab: featurecomp}

\end{table}

\paragraph{Training details}
We train our model in two stages. In the first stage, we train using a volume rendering pipeline and loss $\mathcal{L}_{1}$ defined as~\refEq{eq:l1}. Specifically, we render random patches in size 128×128 cropped from the original 512×512 resolution images during training, which conserves GPU memory while increasing local resolution. However, we observed that initiating training with too large a scaling factor can impede model convergence. Therefore, we progressively increase the original resolution linearly with epochs, from 192 to 512. Additionally, we increase the likelihood of selecting crops covered by the foreground mask. This stage takes 4 days with batch size 32. We
used the Adam optimizer with a learning rate of 4e-4, weight decay of 0.05, and betas of (0.9, 0.95). The learning rate is cosine annealed to 0 during the training.  In the second stage, we train the model with Flexicube pipeline with full resolution using the loss defined as~\refEq{eq:l2}. This stage takes 2 days with batch size 32. We use the same optimization settings as in the first stage, merely adjusting the learning rate to 1e-5. All training is conducted on 16 NVIDIA A6000 48GB GPUs.

\subsection{Comparison}
We compare our method with the previous state-of-the-art feed-forward generation methods, including One-2-3-45~\cite{liu2023one}, LRM~\cite{hong2023lrm}, CRM~\cite{crm} and LGM~\cite{LGM}. We compare the different features of various methods in~\refTab{tab: featurecomp}. We evaluate our method on the Google Scanned Object (GSO) dataset~\cite{downs2022google}. To quantitatively evaluate the image quality synthesized by our approach, we adopt three standard metrics: Peak Signal-to-Noise Ratio (PSNR), Structural Similarity Index (SSIM)~\cite{wang2004image}, and Learned Perceptual Image Patch Similarity (LPIPS)~\cite{zhang2018unreasonable}. Regarding geometric quality, we report the Chamfer Distance (CD) and Volume Intersection over Union (IoU) as metrics. The quantitative results, as shown in~\refTab{tab:delit compare}, demonstrate that our method outperforms others in both color and geometry. Additionally, we show a qualitative comparison of various methods in the~\refFig{fig:compare_image}. As shown in the figure, these methods have all achieved good results, but some of them struggle with certain detailed aspects. One2345~\cite{liu2023one} is limited by the discrete SDF volume grid representation, resulting in blurred textures and geometry. LRM~\cite{hong2023lrm}, lacking multi-view images as constraints, produces 3D models with inconsistent back and front colors and suffers from multi-face Janus problem~\cite{shi2023mvdream}. LGM~\cite{LGM}, using 3DGS as its representation, tends to produce blurred details in areas with dense geometry, such as leaf clusters. CRM~\cite{crm} exhibits geometric discontinuities in the frog's hands and the flower's petioles. However, our method successfully generates 3D assets that are well-aligned with the input conditions, featuring smooth and intact geometric structures.

\begin{table}
    \setlength{\tabcolsep}{1.0mm}
    \centering
    \caption{Quantitative results demonstrate that the color and geometric quality of the 3D assets generated by our method outperforms other methods.}
    \vspace{0.2cm}
  \begin{tabular}{l|ccccc}
    \Xhline{0.8pt}
         & PSNR ↑  & SSIM ↑   & LPIPS ↓    & CD ↓  & IoU ↑        \\ \hline
    One-2-3-45 & 18.93     & 0.779    & 0.166       & 0.0614  & 0.4126               \\
    CRM & 21.59       & 0.864   & 0.159            & 0.0335  & 0.4213               \\
    OpenLRM   & 20.29   & 0.829    & 0.186              & 0.0482   & 0.3731                  \\
    LGM        & 20.05   & 0.798     & 0.176        & 0.0417   & 0.4331                 \\
    Ours         & \textbf{22.52} & \textbf{0.873} & \textbf{0.143}  & \textbf{0.0241}  & \textbf{0.4361}     \\ 
    
    \Xhline{0.8pt}
        
    \end{tabular}
    \vspace{-0.3cm}
    \label{tab:delit compare}
\end{table}

\subsection{Ablation Study}

\paragraph{The effect of illumination decoupled texture.}
\label{sec:ab_illu_decopled}
As shown in~\refFig{fig:ablation_decounpled}, We generate 3d asserts using LDM, applying both illumination-decoupled albedo texture and original RGB texture. Then, we relight the generated results in new scenes. It can be observed that the area highlighted by the red arrow contains excessive shadows, which is incorrect. Without proper illumination decomposition, the shadows baked into the RGB texture from the original lighting combine with the shadows of the new scene, leading to inaccurate results. Therefore, the generation of illumination-decoupled 3D assets plays a crucial role in the usability of downstream applications.

\begin{figure}[h] 
\centering 
\centering
    \includegraphics[width=\linewidth]{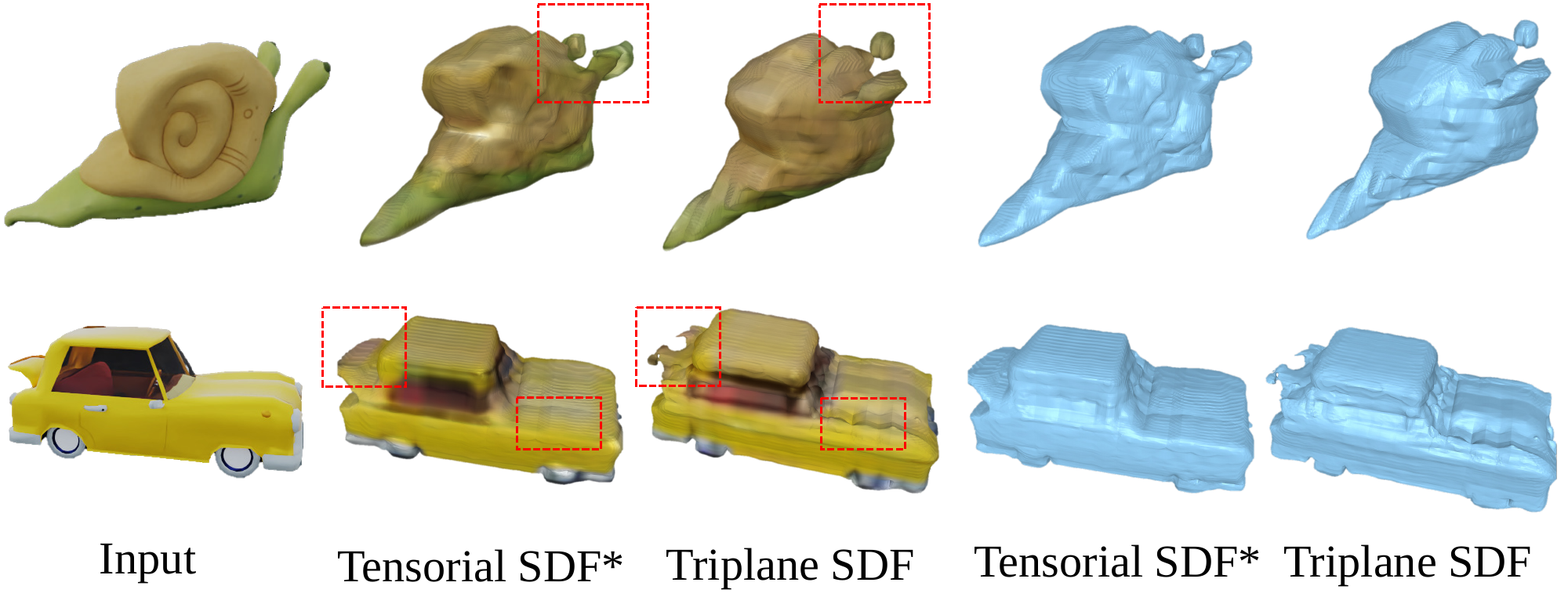}
    \caption{Qualitative comparison results of Tensorial SDF* and Triplane SDF. Note, for a fair comparison, the Tensorial SDF* used here is a scaled small model aligned with Triplane SDF, not the full model.}
    \label{fig:ablation_reps_qul}
    \vspace{-0.5cm}
\end{figure}

\paragraph{Comparisons across different object representation.} 
We conducted ablation studies on various object representations to validate their expressiveness and training convergence speed. Note that due to the substantial training costs associated with our final model, the subsequent analytical experiments employ a significantly reduced version of the LDM model. Inspired by previous works~\cite{hong2023lrm,instant3d}, we reduced the training dataset to a subset containing 10k objects to accelerate model convergence. We conducted training without image cropping using volume rendering, rendering the images at a fixed resolution of 128$\times$128. Each model was trained on 8 NVIDIA A6000 GPUs over 40 epochs. The quantitative results, as shown in~\refFig{fig:ablation_reps}, indicate that compared to the representation of Triplane SDF, the Tensorial SDF representation achieves better quality under the same number of epochs of training and converges faster. Furthermore, it can be observed that constructing objects as Tensorial SDF rather than Tensorial NeRF also aids in convergence. We believe this may be because the SDF representation introduces stronger smoothness constraints, reducing geometric disintegration. Additionally, qualitative comparisons of results, as illustrated in~\refFig{fig:ablation_reps_qul}, demonstrate that Tensorial SDF exhibits superior geometric details with smoother surfaces (e.g., the car hood).

% \begin{figure}[h] 
% \centering 
% \includegraphics[width=0.5\linewidth]{figures/tensor_ablation.png}
% \caption{Comparing model training performance across different object representations, Tensorial SDF emerges as the outstanding performer.}

% \label{fig:ablation_reps} 
% \end{figure}

% \begin{figure}[ht] 
% \centering 
% \includegraphics[width=\linewidth]{figures/tensor_vs_triplane.pdf}
% \caption{Qualitative comparison results of Tensorial SDF* and Triplane SDF. Note, for a fair comparison, the Tensorial SDF* used here is a scaled small model aligned with Triplane SDF, not the full model.}

% \label{fig:ablation_reps_qul} 
% \end{figure}

\begin{figure}[h] 
    \centering
    \includegraphics[width=\linewidth]{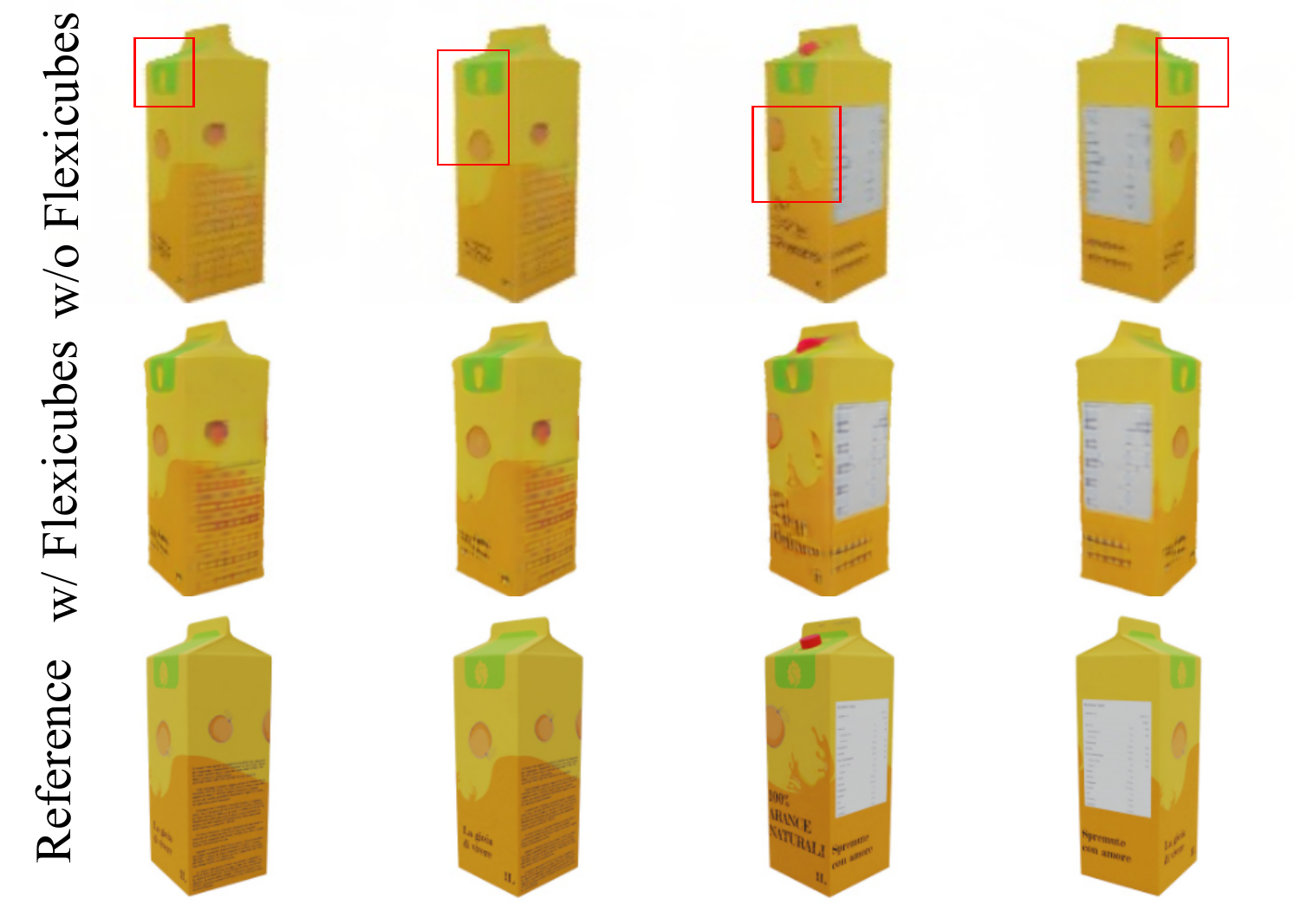}
    \caption{Ablation study shows that the Flexicubes layer helps improve the texture details of generated objects.}
    \label{fig:ablation_flexlayer}
    \vspace{-0.5cm}
\end{figure}

\paragraph{The effect of Flexicubes layer}
We conducted ablation experiments to verify the impact of introducing the Flexicube layer for model fine-tuning. As shown in ~\refFig{fig:ablation_flexlayer} and ~\refTab{fig:ablation_vws_flx}, training with the Flexicubes layer effectively enhances the texture clarity, thanks to the utilization of higher resolution during training. In addition, introducing depth constraints leads to smoother geometric surfaces.

% \begin{figure}[ht] 
% \centering 
% \includegraphics[width=\linewidth]{figures/flex_layer.pdf}
% \caption{Ablation study shows that the Flexicubes layer helps improve the texture details of generated objects.}.  

% \label{fig:ablation_flex} 
% \end{figure}

\begin{table}
  \setlength{\tabcolsep}{2.8mm}
  \centering
  \caption{Quantitative effect of the Flexlayer, evaluate the performance of the fine-tuned model with varying numbers of input views. More discussion can be find in ~\refSec{sec:ab_views}. }
  \vspace{0.2cm}
    \begin{tabular}{c|c|c|c}
        \Xhline{0.8pt}
        Method  &PSNR↑ &SSIM↑ & LPIPS↓    \\
        \Xhline{0.8pt}
        w/ Flex     &23.21   &0.869   & 0.131   \\
        w/o Flex      & 22.76 & 0.853  & 0.145    \\
        w/ view finetune      & 22.89 & 0.856  & 0.149  \\
        w/o view finetune      & 21.63 & 0.786   & 0.157    \\
        \Xhline{0.8pt}
    \end{tabular}
    \vspace{-0.3cm}
    
    \label{fig:ablation_vws_flx}
\end{table}

% \begin{figure}[htbp]
%   \centering
%   \scriptsize
%   \begin{minipage}[c]{0.55\textwidth}
%     \includegraphics[width=\textwidth]{figures/tensor_vs_triplane.pdf}
%     \caption{Qualitative comparison results of Tensorial SDF* and Triplane SDF. Note, for a fair comparison, the Tensorial SDF* used here is a scaled small model aligned with Triplane SDF, not the full model.}
%     \label{fig:ablation_reps_qul}
%   \end{minipage}
%   \hfill % 这个命令在两个子图像之间添加一些水平空间
%   \begin{minipage}[c]{0.42\textwidth}
%   \centering
%     \begin{tabular}{c|c|c|c}
%         \Xhline{0.8pt}
%         Method  &PSNR↑ &SSIM↑ & LPIPS↓    \\
%         \Xhline{0.8pt}
%         w/ Flex     &23.21   &0.869   & 0.131   \\
%         w/o Flex      & 22.76 & 0.853  & 0.145    \\
%         w/ view finetune      & 22.89 & 0.856  & 0.149  \\
%         w/o view finetune      & 21.63 & 0.786   & 0.157    \\
%         \Xhline{0.8pt}
%     \end{tabular}
%     \caption{Quantitative effect of the Flexlayer, evaluate the performance of the fine-tuned model with varying numbers of input views. More discussion can be find in ~\refSec{sec:ab_views}. }
%     \label{fig:ablation_vws_flx}
%   \end{minipage}
%   % \caption{A}
%   \label{fig:images}
% \end{figure}

% \input{tables/ablations_views_flx}

% \begin{figure}[ht] 
% \centering 
% \includegraphics[width=\linewidth]{figures/4view-6view-abl.pdf}
% \caption{Ablation study for transferring the model to a different number of inputs views shows that good results can be achieved with simple fine-tuning.}

% \label{fig:ablation_views} 
% \end{figure}
\begin{figure*}[h] 
\centering   
    \includegraphics[width=\linewidth]{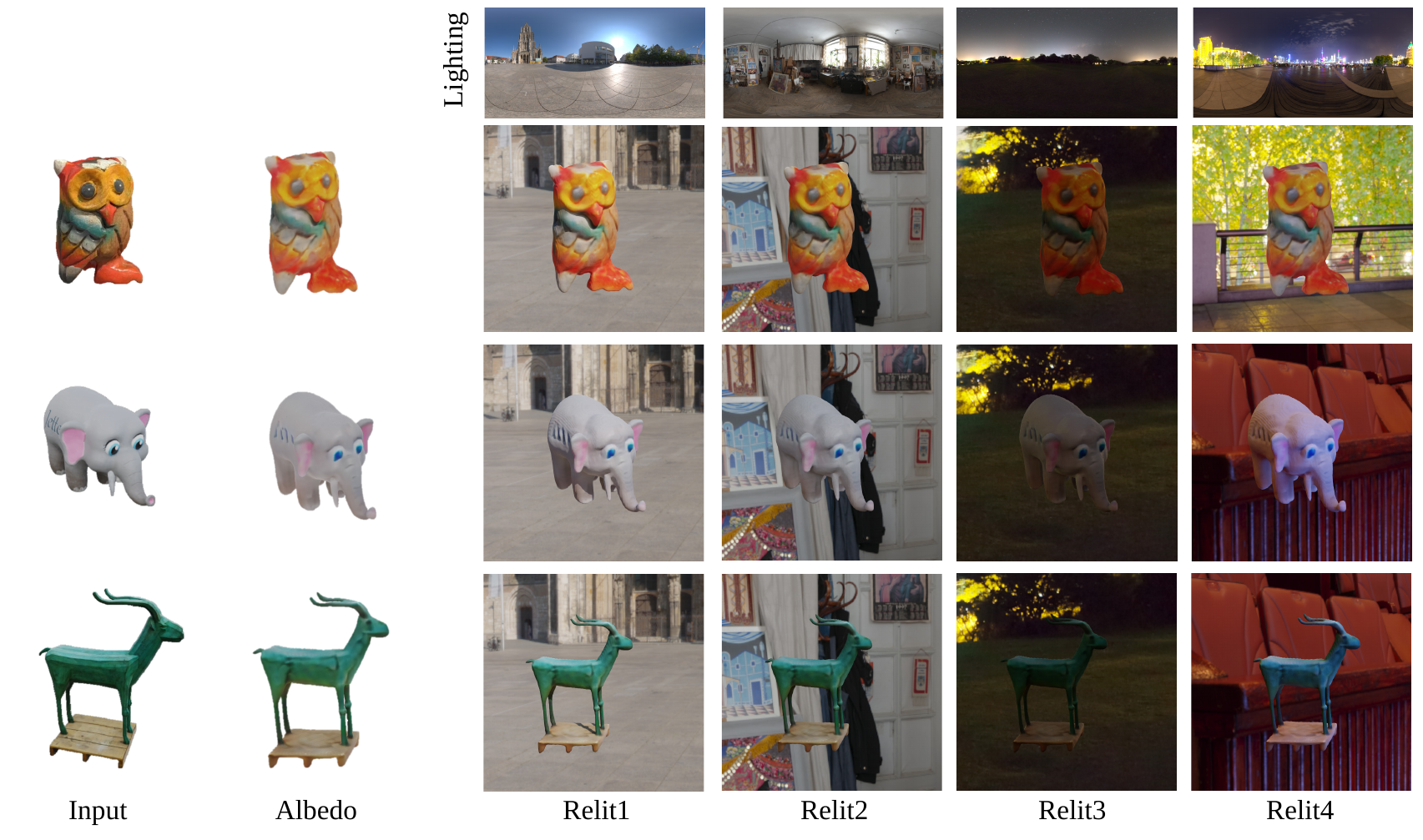}
    \caption{Relighting results of 3d assets generated by our framework.}
    \label{fig:relighting}
    \vspace{-0.3cm}
\end{figure*}

\begin{figure}[h] 
    \centering
    \vspace{-0.2cm}
    \includegraphics[width=\linewidth]{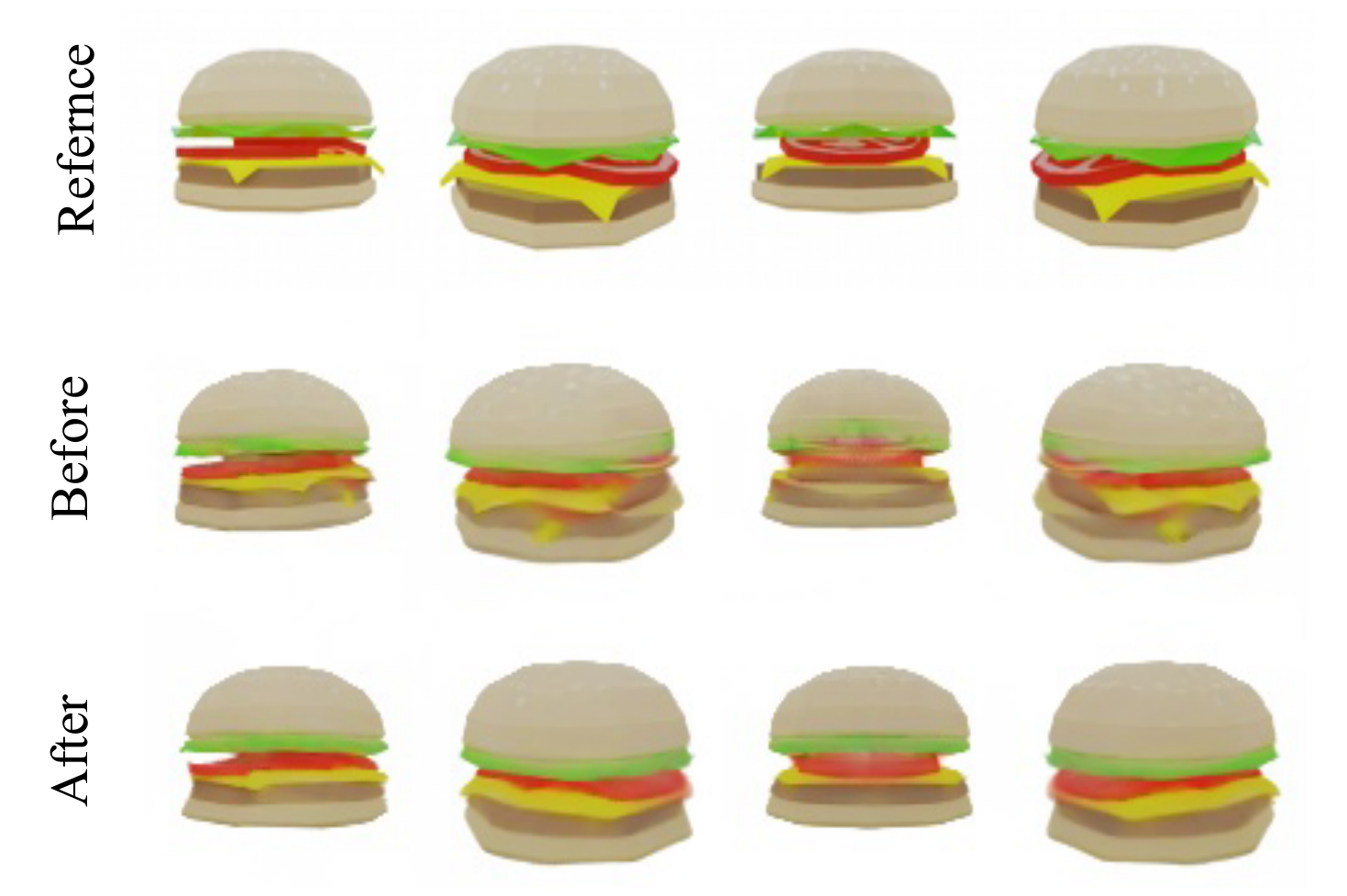}
    \caption{blation study for transferring the model to a different number of inputs views shows that good results can be achieved with simple fine-tuning.}
    \label{fig:ablation_views}
    \vspace{-0.5cm}
\end{figure}

\paragraph{Comparisons across different diffusion methods}
\label{sec:ab_views}
As most text and image-conditioned multi-view diffusion models generate 4 different views, our framework is trained under the assumption of having 4 input views. We evaluated the performance of our framework when switching between multi-view images generated by different diffusion models. Specifically, we try to switch the input of our model to the image predicted by popular multi-view diffusion models Zero123plus~\cite{shi2023zero123++}. The experimental results, as shown in~\refFig{fig:ablation_views} and ~\refTab{fig:ablation_vws_flx}, indicate that when we directly employ a model trained with 4 views to reconstruct 6 views as input, it does not perform well and produces some misaligned results. However, after fine-tuning our model with 6 views input for only 10k iterations, it is able to generate reliable results. This experiment demonstrates that our framework exhibits good generalization across different multi-view diffusion results.

% \begin{figure}[htbp]
%   \centering
%   \begin{minipage}[b]{0.48\textwidth}
%     \includegraphics[width=\textwidth]{figures/flex_layer.pdf}
%     \caption{Ablation study shows that the Flexicubes layer helps improve the texture details of generated objects.}
%     \label{fig:ablation_flexlayer}
%   \end{minipage}
%   \hfill % 这个命令在两个子图像之间添加一些水平空间
%   \begin{minipage}[b]{0.48\textwidth}
%     \includegraphics[width=\textwidth]{figures/4view-6view-abl.pdf}
%     \caption{blation study for transferring the model to a different number of inputs views shows that good results can be achieved with simple fine-tuning.}
%     \label{fig:ablation_views}
%   \end{minipage}
%   % \caption{A}
%   \label{fig:images}
% \end{figure}

\begin{figure*}[h] 
\centering  
    \includegraphics[width=\linewidth]{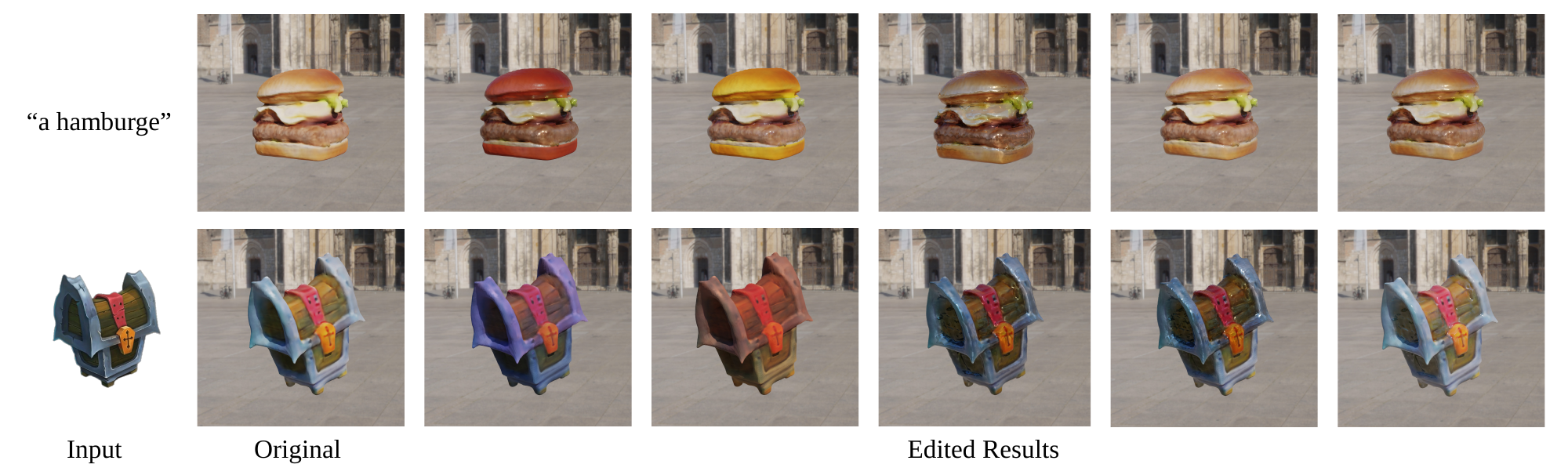}
    \caption{We applied different material textures to the generated 3D assets and modified the albedo color to produce some interesting results, such as a metallic hamburger.}
    \label{fig:editing}
    \vspace{-0.3cm}
\end{figure*}

\section{Application}
As our method generates 3D assets with illumination-decomposed texture maps, it easily supports applications such as relighting and material editing. As shown in~\refFig{fig:relighting}, we relight the 3D assets in different new scenes, producing convincing synthesized images. Additionally, we apply PBR material textures to the 3D assets and edit their metallicity, roughness, and albedo properties, with the resulting images shown in~\refFig{fig:editing}.

\begin{figure*} [h]
\centering 
\includegraphics[width=\linewidth]{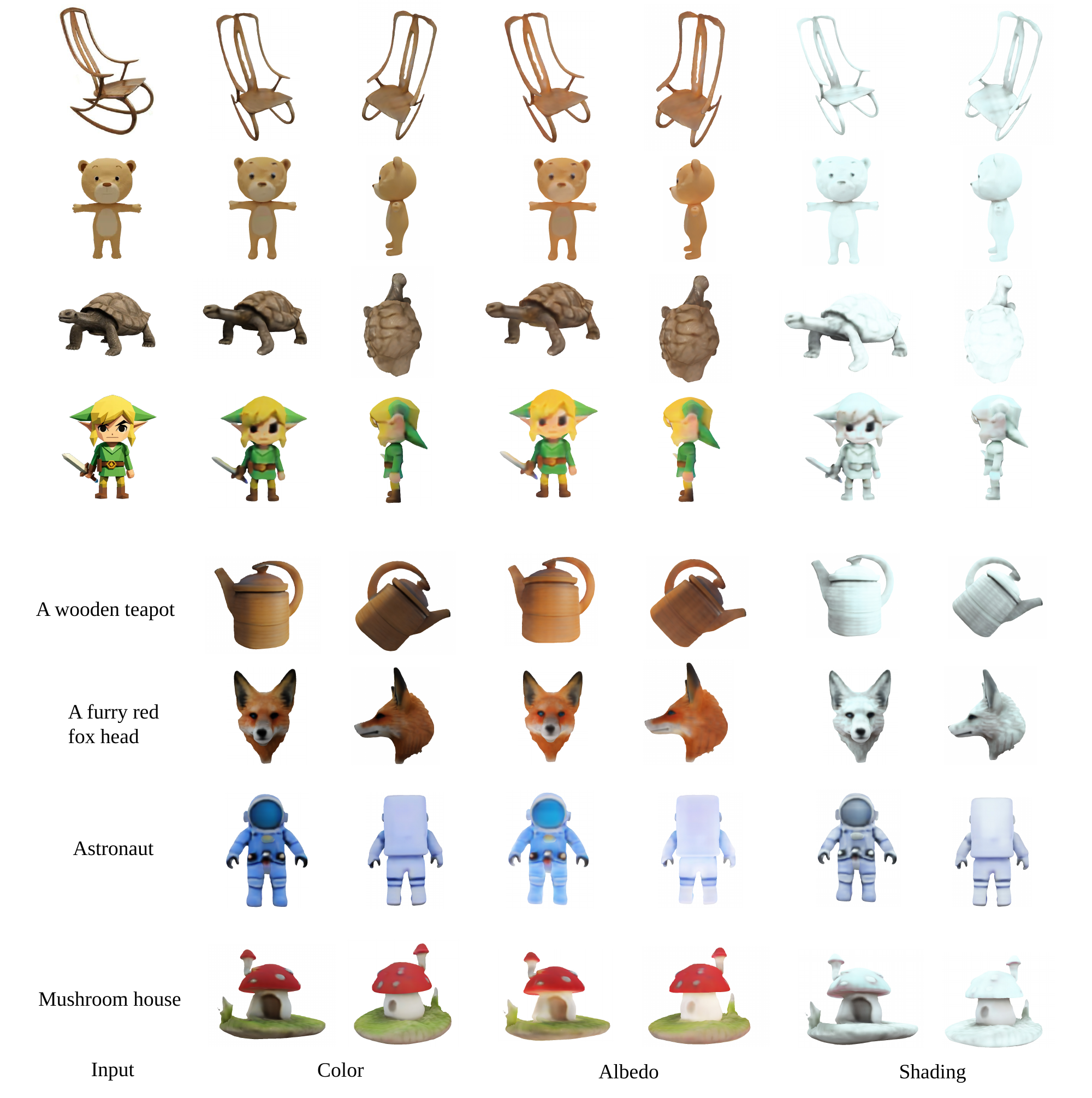}

\caption{More predicted results for text-to-3D generation task.}

\label{fig:more_img23d} 
\end{figure*}

\section{Limitations}
\label{sec:discuss}

While our model can generate high-quality illumination-decoupled 3D assets, there are still some limitations. Like other transformer-based approaches~\cite{instant3d,hong2023lrm}, the size of the tensorial SDF tokens produced by our model is capped at 64x64, constraining the resolution of the final 3D assets. In addition, our illumination-decoupled module is not designed to handle complex materials, such as translucent surfaces, due to our simplified rendering composition. Finally, the quality of our reconstructed 3D models is influenced by the inconsistencies in the multi-view diffusion. We reserve these challenges for future exploration. Further discussion can be found in the appendix~\refSec{sec:lim_appendix}.

\section{Conclusion}
\label{sec:conclusion}
This paper introduces a novel feed-forward framework capable of reconstructing 3D meshes with illumination-decoupled textures from a single image input or text prompt in 10 seconds. Our method utilizes a conditional multi-view diffusion model to generate consistent four-view images and lifts them to 3D using a transformer-based large reconstruction model. We propose to generate objects represented as tensorial SDF field, which is more expressive and can accelerate model convergence compared to previous tri-plane representations. In addition, we decompose the rendering color into albedo color and shading color, which enables the generated 3D assets to be easily relit and edited for material properties. We believe that the ability to produce relightable 3D assets is very important for downstream applications.

\section*{Appendices}

\appendix

\section{Detail for Network architecture}
We use the DINOv2-ViT-B/14~\cite{dino} as our image encoder, a transformer-based model, which has 12 layers and the hidden dimension of the transformer is 768. Regarding camera features, we flatten the extrinsic parameters of each view into 16 dimensions and encode them into a 1024-dimensional vector using a 2-layer MLP. After encoding, each image yields 257 image feature tokens, including the [CLS] token. All tokens from these views are concatenated together to obtain a total of N$\times$257 tokens, where N is the number of input images. These image feature tokens serve as condition features in the subsequent generation process.

Next, we use a transformer-based tensorial object reconstructor to predict a tensorial SDF representation from a sequence of learnable tokens with a size of (3$\times$32$\times$32+3$\times$32)$\times$1024, where 3$\times$32$\times$32 and 3$\times$32 are the number token numbers align with $V^m_k$ and ${M}_k^{\tilde{m}}$ represent the tensor matrix factors and vector of their corresponding spatial axes. And 1024 is the hidden dimension of the transformer decoder. After being decoded by the transformer, conducted through cross attention from image features, we obtain the same number of tensor tokens.
Next, following LRM\cite{hong2023lrm}, we use a de-convolution layer and an MLP layer to upscale tensor matrix factors from 3$\times$(32$\times$32)$\times$1024 to 3$\times$(64$\times$64)$\times$40, and tensor vector from 3$\times$32$\times$1024 to 3$\times$64$\times$40. 

Finally, these tensor tokens are reshaped into a tensorial SDF representation, from which we can compute the feature vector of any point with a dimension of 120. Then, for the first stage of volume rendering training, we use a 4-layer MLP with 64 hidden dimensions to decode a 1-dim SDF value from this feature. We use another 4-layer MLP with 64 hidden dimensions to decode a 6-dim color value from this feature, where the first three dimensions are considered the albedo color and the last three dimensions are considered the shading color. Finally, for the second stage of Flexcubes layer~\cite{flexible} training, we use a 2-layer MLP with 64 hidden dimensions to decode an 8-dim weights value from this feature. We use another 2-layer MLP with 64 hidden dimensions to decode a 1-dim deformation value from this feature.

\section{More results for 3D generation task}

As shown in~\refFig{fig:more_img23d}, we present more 3D results generated from text prompts or single image inputs. Our method is capable of producing good results for both real-world images and unreal images.

\section{Failure Cases and Limitations}
\label{sec:lim_appendix}
\begin{figure*}[t]
\centering
\includegraphics[width=\linewidth]{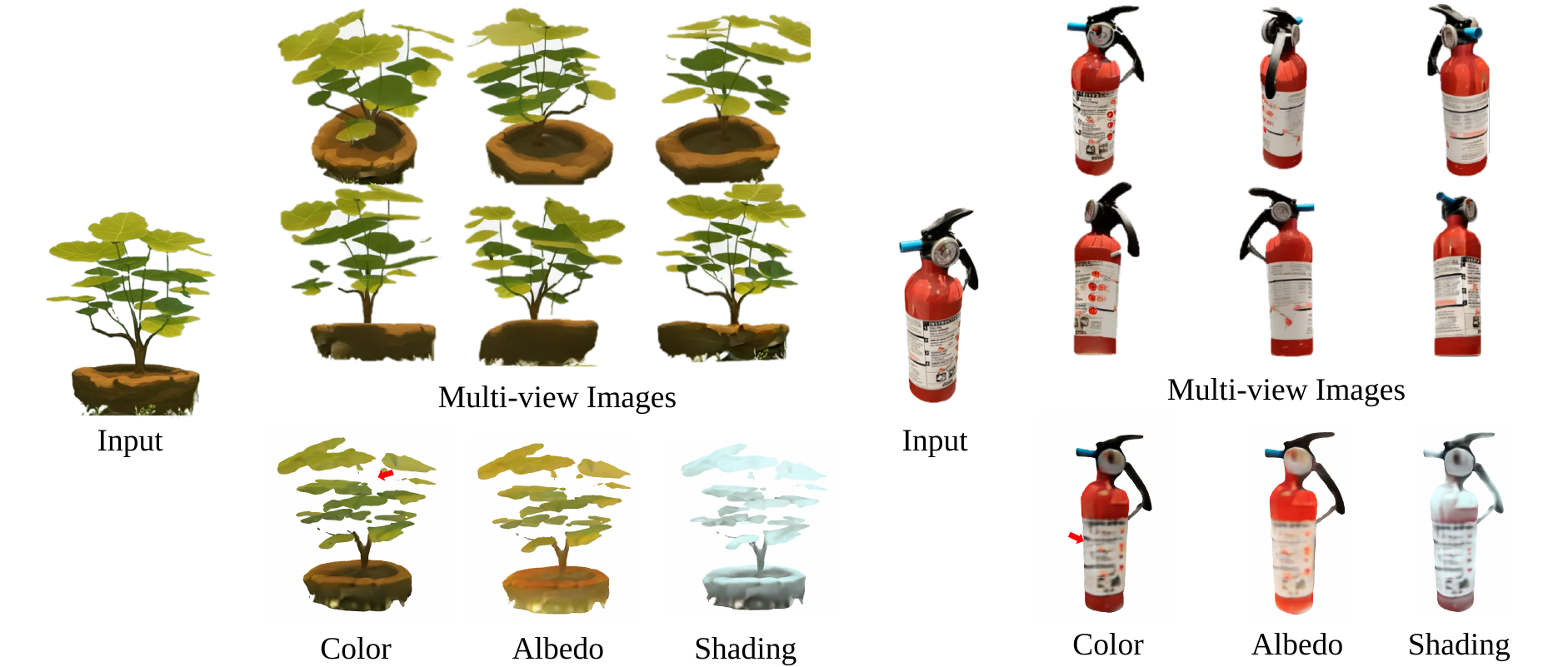}

\caption{Some failure cases in the predictions reveal the limitations of our method.}

\label{fig:failure_cases} 
\end{figure*}

As discussed in the main text, our framework can produce convincing results from a single image or text prompt input, but there are still certain instances where it fails, as shown in~\refFig{fig:failure_cases}. In the left side of~\refFig{fig:failure_cases}, we input an image of a potted plant to predict its 3D assets. The overall result meets expectations, with the plant's leaves and base being well generated, but the thin stems fail to generate properly. There are multiple reasons for this issue. On the one hand, the size of the tensorial SDF tokens produced by our model is capped at 64x64, which may result in the loss of fine geometric details. On the other hand, the multi-view images generated by the multi-view diffusion may be inconsistent, causing thin stems to misalign across different views, leading to generation failures. On the right side of~\refFig{fig:failure_cases}, we can see that due to misalignment in the multi-view images, the text on the predicted fire extinguisher appears blurred. The misaligned highlights also caused errors in the material prediction, with the highlights being incorrectly predicted as part of the albedo. Additionally, the prediction of the shading color for the metallic and lacquer materials on the fire extinguisher's surface is inaccurate. While part of the red metallic surface is correctly predicted in the shading color, some of the black text is mistakenly predicted as part of the shading color as well.

%-------------------------------------------------------------------------

{\small
\bibliographystyle{cvm}
\bibliography{cvmbib}
}

% \section{More results for 3D generation task}

% As shown in~\refFig{fig:more_img23d}, we present more 3D results generated from text prompts or single image inputs. Our method is capable of producing good results for both real-world images and unreal images.

\end{document}